\documentclass[12pt,preprint]{aastex}

\shorttitle{G272.2-3.2 in X-rays}
\shortauthors{McEntaffer et al.}

\begin{document}

\title{X-ray Emission from the Galactic Supernova Remnant G272.2-3.2}

\author{R. L. McEntaffer, N. Grieves, \& T. Brantseg}
\affil{Department of Physics and Astronomy, University of Iowa, Iowa City, IA 52242}
\email{randall-mcentaffer@uiowa.edu}

\begin{abstract}
We present analysis of \textit{Chandra X-ray Observatory} data detailing a galactic supernova remnant, G272.2-3.2.  A clear shell of emission has been resolved for the first time as a series of filaments and knots around the entire rim of the remnant.  Spectral analysis of these features show that they are consistent with shock heating of interstellar material in a clumpy medium.  Spatially separated from this shell we see a central diffuse region dominated by harder, hotter emission.  Spatial spectroscopy shows a clear enhancement of metals consistent with a Type Ia explosion, namely S, Si, and Fe.  We find no clear evidence for a compact object or pulsar wind nebula and argue for a Type Ia origin.  Consideration of the ionization timescales suggest an age of 8700 years for G272.2-3.2.
\end{abstract}

\section{Introduction}\label{intro}
The galactic supernova remnant (SNR) G272.2-3.2 was originally identified in the all sky survey of \textit{ROSAT} \citep{GreinerEgger}.  Using multiple techniques, \citet{Greiner1994} find a possible age range of 400--4000 years and a distance of $1.8^{+1.4}_{-0.8}$ kpc for this object.  The X-ray emission exhibits circular morphology with a size of 15.2 arcminutes.  The authors find a spectrum that is thermal in origin possibly arising from a reverse shock reheating ejecta from the supernova explosion or from cloud evaporation in the hot, shocked interior.

Faint radio emission was detected from the source using the 64 m Parkes radio telescope, the Australia Telescope Compact Array (ATCA), and the Molonglo Observatory Synthesis Telescope (MOST) \citep{Duncan1997}.  Diffuse radio emission fills a circular region of $\sim15$ arcminutes agreeing with X-ray data.  In this region there are four radio ``blobs'', one of which is near the center of the remnant.  Three of these have strong spatial correlations to optical filaments studied by \citet{Winkler1993}.  The filaments have [\ion{S}{2}]/H$\alpha$ ratios consistent with an SNR.  The radio/optical correlation suggests shock interaction in a clumpy circumstellar environment.  The radio fluxes indicate non-thermal emission with a spectral index of $\alpha\sim0.55$ consistent with a shell type remnant.  The authors find no evidence for thermal emission or spectral indices expected from a pulsar wind nebula (PWN).  Furthermore, the radio data do not exhibit polarized emission given the 1 arcminute beam size suggesting a turbulent magnetic field, at least down to this scale.  Given the presence of optical filaments and a tentative anti-correlation between the X-ray and radio features \citet{Duncan1997} argue for an older age of $\sim6000$ years.

\citet{Harrus} performed an X-ray study using data from the \textit{ASCA} and \textit{ROSAT} observatories.  They classify G272.2-3.2 as a thermal composite SNR due to its centrally peaked thermal emission.  They find evidence for areas of enhanced emission, especially in the west, but do not detect the presence of a shell.  A spectrum of the entire remnant can be characterized by a nonequilibrium plasma with a temperature of $kT\sim0.73$ keV.  To explain the centrally peaked morphology the authors consider the possibility of clouds evaporating in a hot, shocked medium \citep{WhiteLong} versus temperature equilibration via thermal conduction in a Sedov type remnant \citep{Cox1999}.  They conclude that the former is preferred in addition to the possibility that projection effects along the line of sight may contribute.

More recently, the X-ray emission from G272.2-3.2 has been revisited using \textit{Suzaku}.  \citet{SezerGok} present spectra extracted from three regions -- the entire remnant, the central region (3.8 arcminute radius), and the outer region (annulus between radii of 3.8 and 7.6 arcminutes).  In general, the remnant can be desribed by a nonequilibrium thermal plasma with a temperature of $kT\sim0.77$ keV, consistent with \citet{Harrus}.  The central region is brighter and slightly hotter with enhanced abundances indicative of ejecta while the outer region is slightly cooler with abudances consistent with local interstellar medium (ISM).  Their data also show centrally peaked X-ray emission.  Given the nonequilibrium nature of the plasma they also support cloud evaporation over thermal conduction to explain the morphology.  Finally, when comparing the derived abundances to Type Ia nucleosynthesis yields \citep{Nomoto}, the results are consistent with a Type Ia origin.

In this paper we discuss our analysis of G272.2-3.2 using data from the \textit{Chandra X-ray Observatory}.  Using the higher spatial resolution of \textit{Chandra} we detail the X-ray morphology and spectral properties to a higher degree than previously possible.  The combination of spatial and spectral data provides a keen look into distinct interactions with the local ISM to characterize the X-ray emission in the previously undetected outer shell region of this remnant.  It also allows for a detailed study of the central region to verify the nature of the plasma and the Type Ia designation.

\section{Observations and Data Reduction}
SNR G272.2-3.2 was observed with the Advanced CCD Imaging Spectrometer (ACIS) on \textit{Chandra} for 42 ks and 23 ks on July 26 and 27, 2008. The observations were archived as obsIDs 9147 and 10572.  The entire SNR was imaged on the four front-illuminated ACIS-I chips. These chips have decreased low energy response in comparison to the back-illuminated chips in ACIS-S, but the remnant fits neatly into the $2\times2$ ACIS-I array and lies near the galactic plane suggesting high absorption and highly attenuated soft X-ray flux.  The observation produced imaging with an unbinned spatial resolution of $<\sim1$ arcsecond.

\begin{figure} [htbp]
   \centering
%   \epsscale{1.0}
%   \plotone{f1.eps}
%   \includegraphics[width=6.0in,height=5.76in]{figures/rawxray.pdf}
   \includegraphics[width=6.0in,height=5.76in]{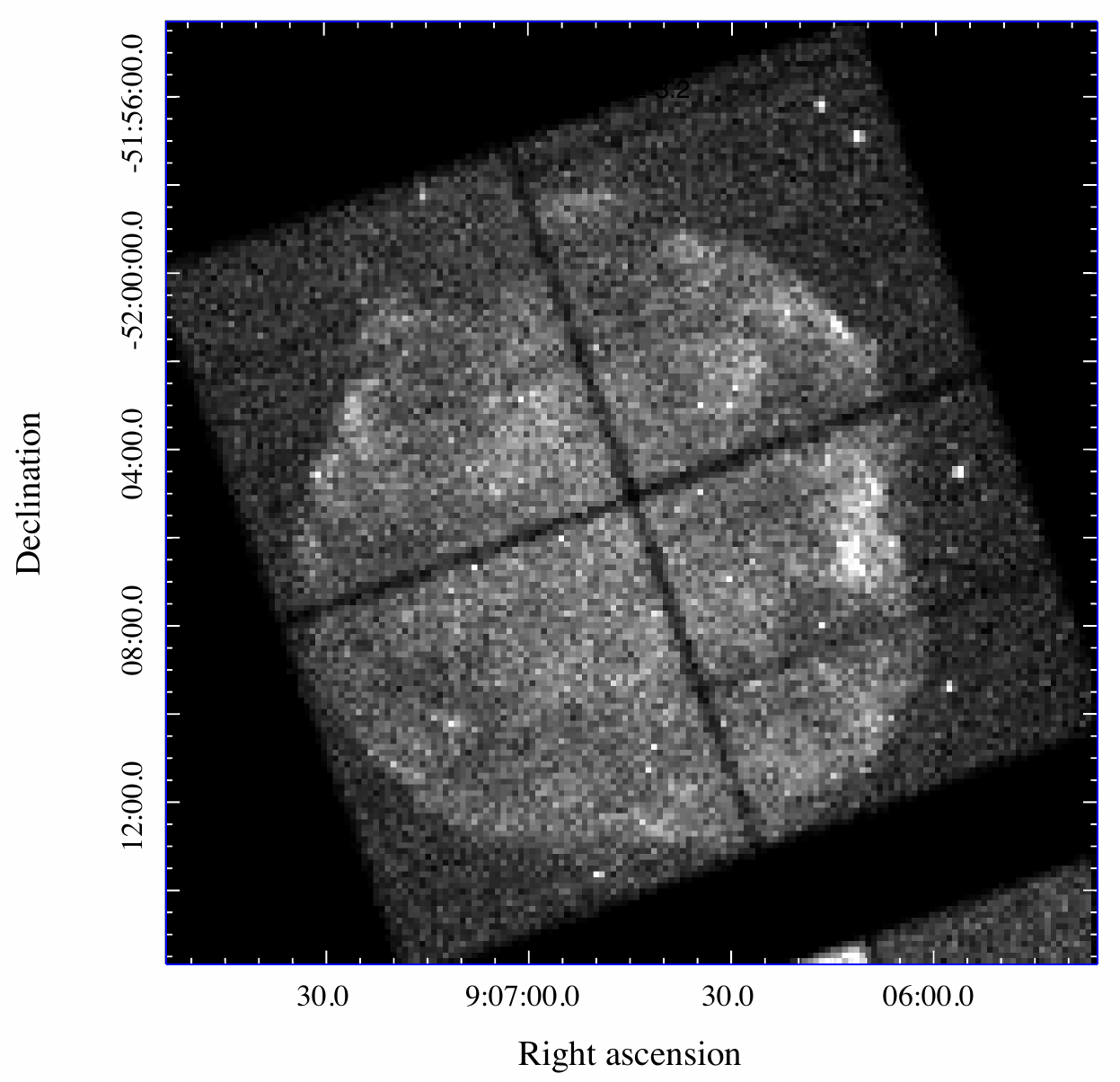}
   \caption{Image of the raw data from the \textit{Chandra} ACIS-I CCDs.  The data are unsmoothed and binned at 16 pixels.  The scale runs from 0--50 counts per bin.}
   \label{rawxray}
\end{figure}

The level 1 data were reprocessed to level 2 with standard processing procedures in the Chandra Interactive Analysis of Observations (CIAO, ver. 4.4; \citet{CIAO2006}) software package with current calibration data from the Chandra Calibration Database (CALDB, v.4.4.2). Good time intervals, charge transfer inefficiency, and time-dependent gain variation were accounted for. The reprocessed raw X-ray data are shown in Figure \ref{rawxray}.  

Infrared images of G272.2-3.2 in the 22 $\mu$m (W4) band from the \textit{WISE} All-Sky Data Release were obtained from the NASA/IPAC Infrared Science Archive (IRSA), corresponding to \textit{WISE} Atlas tiles 1356, 1358, and 1380. These images provide an angular resolution of 12 arcseconds and a minimum point source sensitivity of 6 mJy. In addition to the standard pipeline processing and basic cropping, the images were rescaled to a common intensity scale with the IRAF task \textit{mscimatch}, which photometrically matches the brightness of background stars shared in multiple images. The images were then combined into a mosaic with the IRAF task \textit{mscstack} to produce the final image.

\section{Spectral Extraction and Results}
\subsection{Radial Analysis}
\begin{figure} [htbp]
   \centering
%   \epsscale{1.0}
%   \plotone{f2.eps}
%   \includegraphics[width=6.0in,height=5.76in]{figures/AnnularRegs.pdf}
   \includegraphics[width=6.0in,height=5.76in]{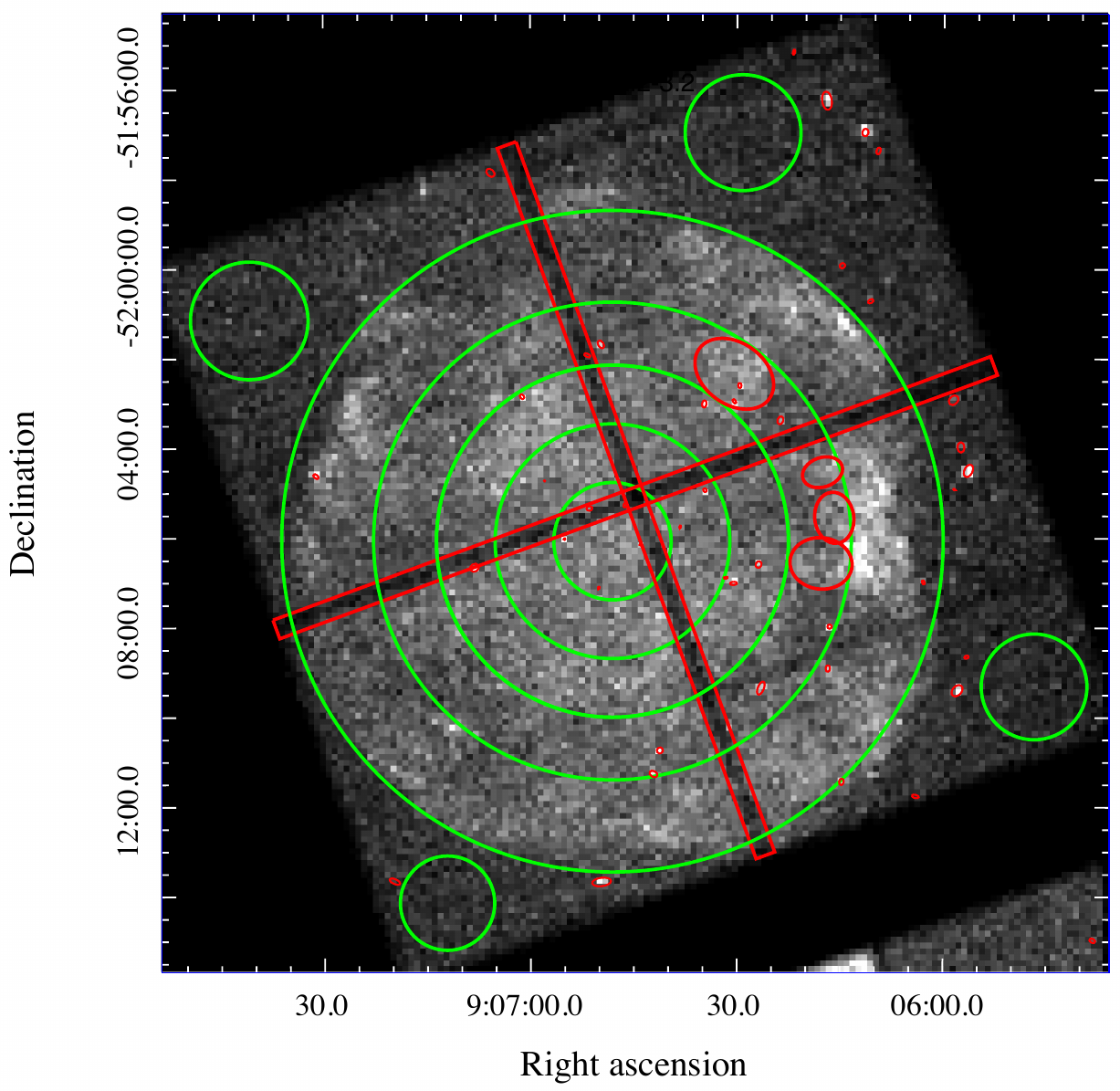}
   \caption{The green annular regions used for radial spectral analysis are shown overlaid on the raw X-ray data.  Regions outlined in red were subtracted from the data prior to spectral fitting.}
   \label{AnnularRegs}
\end{figure}

The image in Figure \ref{rawxray} already displays a much higher degree of detail than observed in any previous study.  There are bright knots, blobs, and filaments along a shell of emission at the rim of the remnant.  Also, there is broad diffuse emission that fills much of the interior.  Furthermore, this interior emission appears distinct and separated from the shell by a darker annulus between them.  To study the apparent radial dependence on morphology we extract annular regions as displayed in Figure \ref{AnnularRegs}.    The inner circle has a radius of 1.3 arcminutes and the radial ranges of the first two annuli are both 1.3 arcminutes as well.  The third annular region attempts to follow the dark separation and has a radial range of 1.4 arcminutes.  The outermost region encompasses the shell and has a radial range of 2.1 arcminutes.  The regions outlined in red are subtracted from the data prior to spectral fitting.  The two narrow rectangles forming the cross eliminate the chip gaps and data near the chip edges.  The multitude of small red ellipses denote the position of point sources.  These regions are determined using \textit{wavdetect} in CIAO \citep{freeman}. This tool finds sources in the data set by correlating the image with ``Mexican Hat'' wavelet functions. The tool then draws an elliptical region around the detected source to a specified number of standard deviations, $\sigma$, where $1\sigma$ is the uncertainty in the point spread function. In the present case, the detection scales used are 2 and 4 pixels, and the region size is $3\sigma$.  Finally, there is a set of four larger red ellipses.  These were subtracted from the annulus corresponding to the dark region between the central emission and the shell emission in order to limit contributions from interactions with local ISM clouds in this possibly intermediate region.  

Spectra for each of the annular regions are extracted using the \textit{specextract} tool in CIAO.  This tool automatically creates ancillary response files and redistribution matrix files for each region.  The extracted spectra were background corrected using four circular regions devoid of emission shown as circles in the corner of each CCD.  The spectral resolution obtained by \textit{Chandra} is $\sim$5--20 ($E/ \Delta E$) over the energy band used, 0.3-8.0 keV.  In our analysis, data below 0.3 keV are not used due to uncertain ACIS calibration while data above 3.0 keV are consistent with zero.  The data are binned to at least 20 counts per bin to allow the use of Gaussian statistics.  To analyze the extracted spectra, we use the CIAO modeling and fitting package, Sherpa \citep{Sherpa}.

Typically, X-ray emission in a diffuse SNR is caused by shock heating of ejecta or ISM, which results in prominent emission lines. Depending on the plasma density and the time since it has been shocked, we expect the emitting material to either be in collisional ionization equilibrium (CIE; if the plasma has high density or was shocked a long time ago) or in nonequilibrium ionization (NEI).  We use the \textit{xsvnei} model for nonequilibrium conditions (\citet{bork94,bork01,Hamilton,liedahl95}.  We replace the default \textit{xsvnei} line list with an augmented list developed by Kazik Borkowski that includes more inner shell processes especially for the Fe-L lines.  For equilibrium conditions we use the \textit{xsvapec} model, which uses an updated version of the ATOMDB code (v2.0.1; \citet{aped1,aped2}) to model the emission spectrum.  We include a second temperature component to these fits if it is determined statistically relevant as determined from an F-test (probabilites $<0.05$ indicate a statistical improvement given the additional component).  Emission line lists in the 0.3--3.0 keV energy range for plasmas with temperatures $kT\sim$0.09--2.0 keV show that the emission is dominated by highly ionized states of C, N, O, Ne, Mg, Si, S, and Fe.  The spectral fits begin with all abundances frozen to solar levels.  A given element or combination of elements is allowed to vary if it significantly improves the fit.  Dielectronic recombination rates are taken from \citet{mazz} with solar abundances from \citet{angr} and cross-sections from \citet{bcmc}.  In addition, we investigate the significance of a contribution from a non-thermal component by including a \textit{xspowerlaw} model.  In the shell this probes for the possibility of synchrotron emission due to swept up magnetic fields while in the interior it tests for the presence of a pulsar wind nebula in the case that the progenitor was a massive star.

\begin{figure} [htbp]
   \centering
%   \epsscale{1.0}
%   \plotone{f3.eps}
%   \includegraphics[width=6.0in,height=4.55in]{figures/annularspecs.pdf}
   \includegraphics[width=6.0in,height=4.55in]{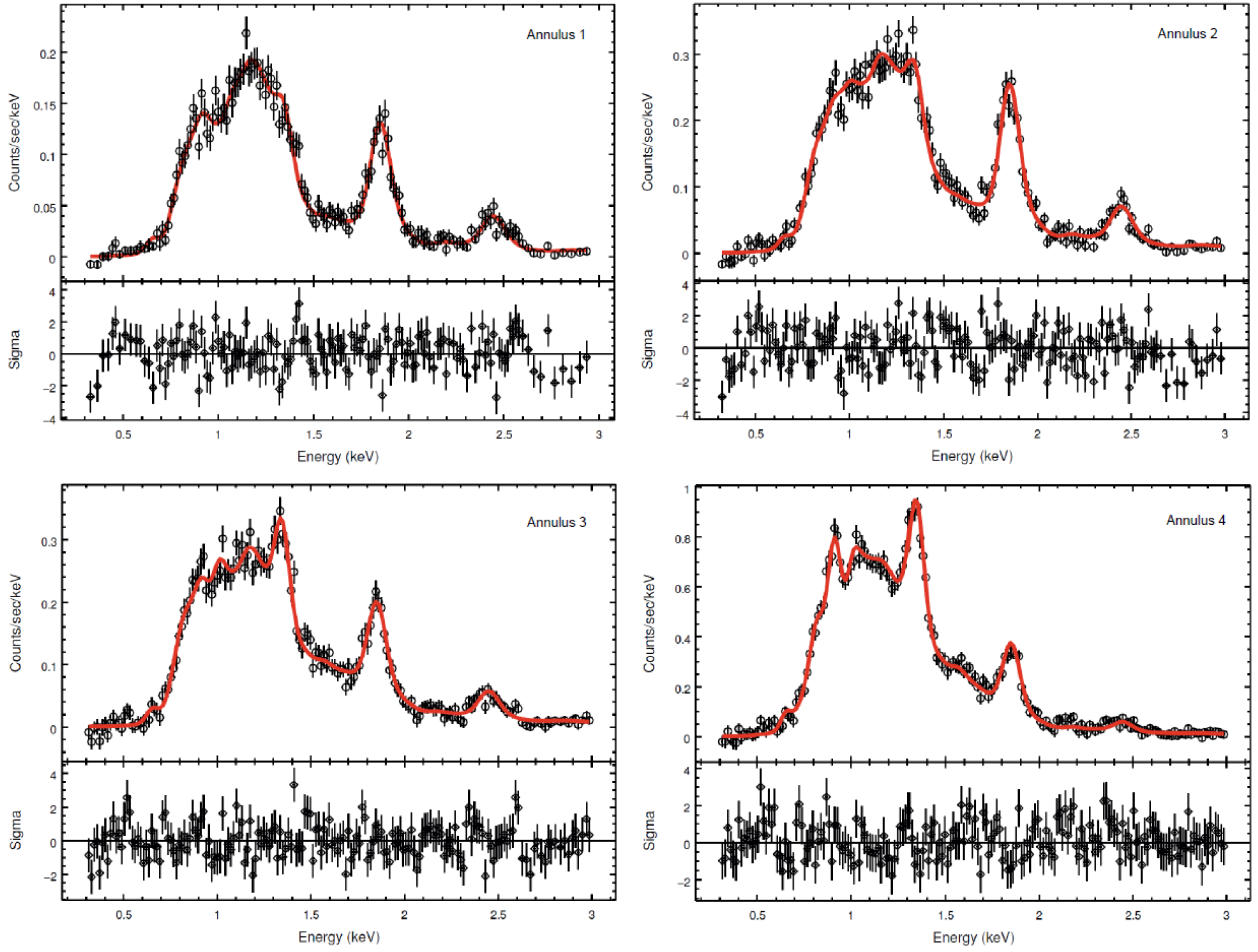}
   \caption{Extracted spectra from annuli 1--4 are shown as the black data points.  Best fit models to these data are shown as red lines.  Fit residuals are displayed below each data set.}
   \label{annularspecs}
\end{figure}

The best fit parameters for the annular regions are shown in Table \ref{table1} and Table \ref{table2}.  The spectra with best fits overlaid are shown in Figure \ref{annularspecs} (the fit to the center circle is very similar to annulus 1 and not shown here).  The center of the remnant and annulus 1 are well described by a high temperature, $kT\sim1.5$ keV, nonequilibrium plasma.  However, as we move into annulus 2, which encompasses the outer limit of the central emission, a contribution from a slightly lower temperature plasma becomes significant.  In fact, this very same component takes over to describe the emission found in annulus 3, which outlines the dark region between the central emission and the shell.  Out in this shell, annulus 4 is best fit by a combination of a mid-temperature nonequilibrium plasma and a cool equilibrium plasma.  The best fit elemental abundances for these regions are shown in Table \ref{table2}.  The inner four annuli all show overabundances of O, Si, S, and Fe relative to solar at the 3--6$\sigma$ level.  In the shell the abundances are typically consistent with solar with an overabundance of O and slightly depleted Mg.  The O abundances are typically loosely constrained and difficult to interpret given the high absorption column and lack of flux at these lines.

\begin{figure} [htbp]
   \centering
%   \epsscale{1.0}
%   \plotone{f4.eps}
%   \includegraphics[width=6.0in,height=5.76in]{figures/tricolor.pdf}
   \includegraphics[width=6.0in,height=5.76in]{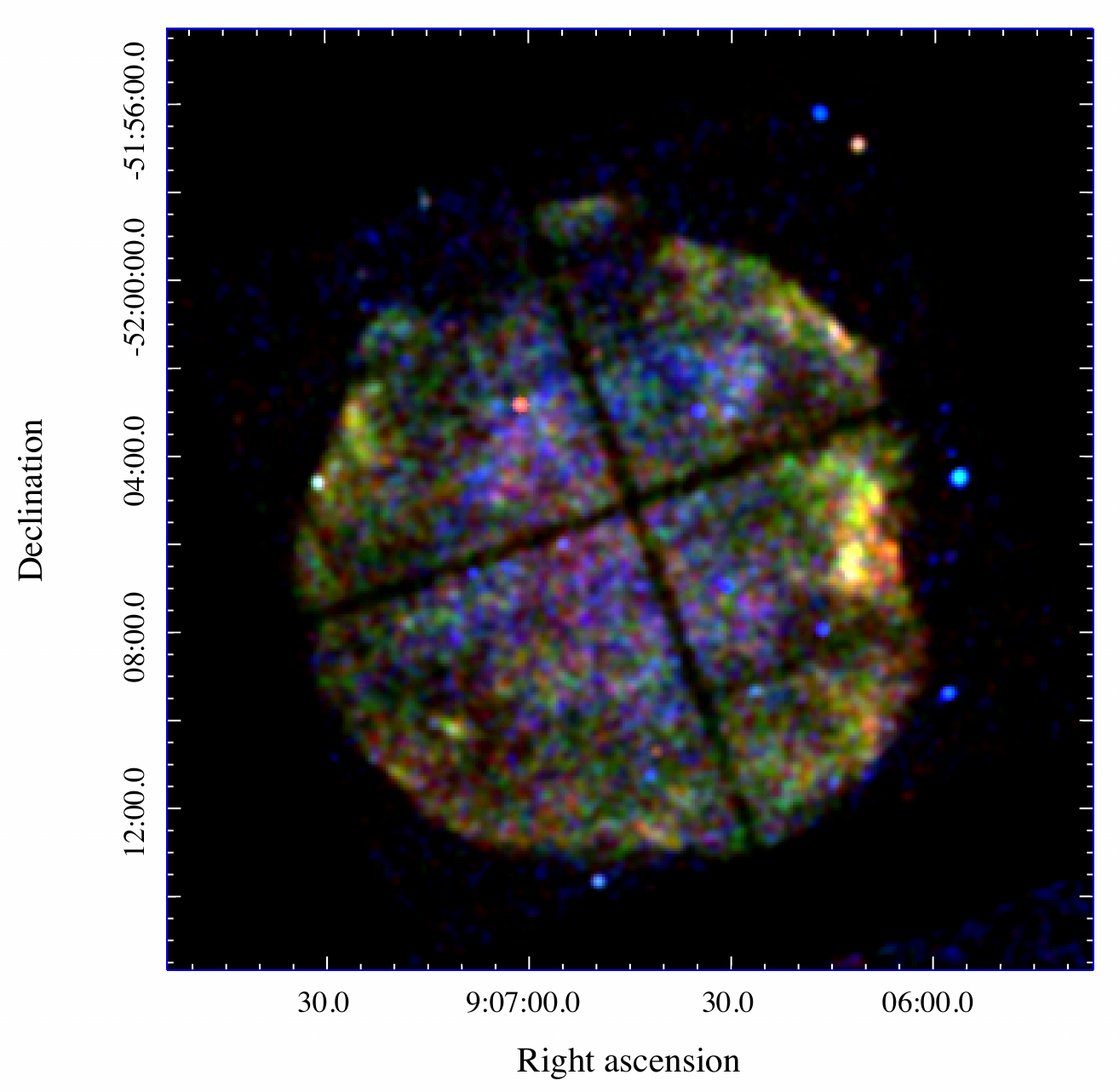}
   \caption{A false-color image of SNR G272.2-3.2 in X-rays.  Red covers the energy range 0.3--1.2 keV with green from 1.2--1.7 keV, and 1.7--8.0 keV for blue.}
   \label{tricolor}
\end{figure}

The temperature is fairly consistent in the center of the remnant and decreases as we move into the outer annuli.  Furthermore, the abundances are higher in annuli 1--3 and decrease rapidly to solar levels going into annulus 5.  These results suggest a hot, ejecta dominated interior with a lower temperature shell of emission arising from shocked ISM.  As shown in Figure \ref{tricolor} we construct a false color RGB image in an attempt to understand this morphology more completely.  The energy ranges covered by the respective bands are 0.3--1.2 keV for red, 1.2--1.7 keV for green, and 1.7--8.0 keV for blue.  As with the spectral fits to the raw data, this beautiful image paints the picture of a hot central region separated from a lower temperature shocked shell.  These data are the first to resolve this shell and we discuss spectral extractions of its features in the following section. 

\subsection{Shell Emission}
The evident shell in Figure \ref{tricolor} is dominated by the lower and middle X-ray energy ranges with most of the higher energy blue emission near the center.  As seen from the radial analysis, the best fit to annulus 5 includes a low temperature equilibrium plasma and a higher temperature nonequilibrium component.  We investigate the plasma conditions in the shell by performing spectral extractions on discrete regions drawn around apparent knots and filaments.  These enhancements could signify shocked clouds in a swept up shell of ISM.  

\begin{figure} [htbp]
   \centering
%   \epsscale{1.0}
%   \plotone{f5.eps}
%   \includegraphics[width=6.0in,height=5.76in]{figures/RGBregions.pdf}
   \includegraphics[width=6.0in,height=5.76in]{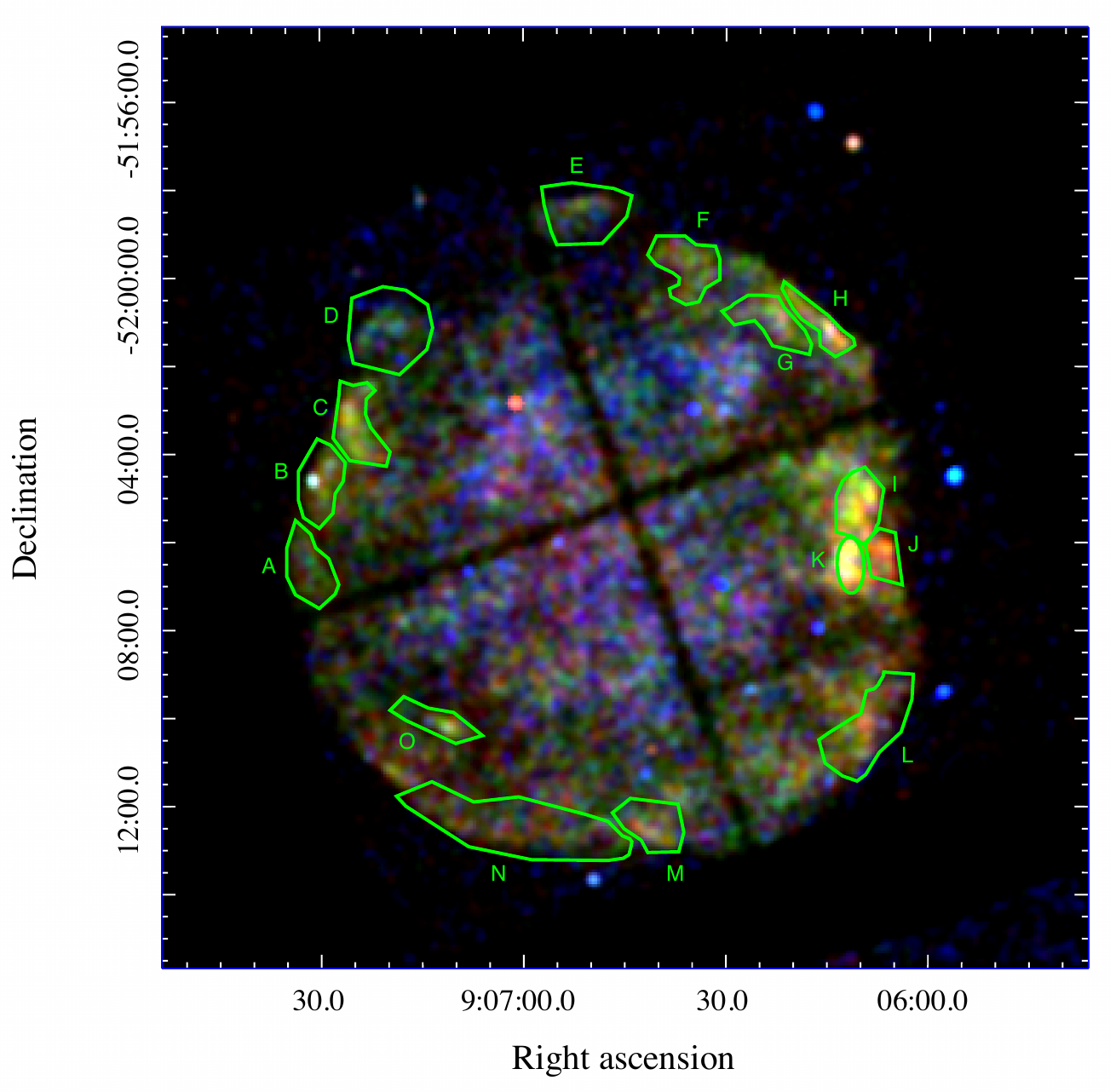}
   \caption{The false-color image of SNR G272.2-3.2 with overlaid extraction regions in green.}
   \label{RGBregions}
\end{figure}

Figure \ref{RGBregions} displays our extraction regions overlaid on the RGB image.  As with the radial analysis we attempt fits containing equilibrium plasmas, nonequilibrium plasmas or some combination thereof, all with variable abundances.  We also test for the presence of a nonthermal powerlaw component, but find no indication of such a component in any of the shell regions.  Table \ref{table3} summarizes the best fit parameters for the extractions.  The fits in the shell echo the fit found from annulus 5 -- they are a mix of low temperature equilibrium plasmas and higher temperature nonequilibrium plasmas.  The exception is region G which has a higher temperature yet is in equilibrium.  A comparable fit was obtained using nonequilibrium conditions and similar parameters, but the \textit{vapec} model was slightly favored in an f-test.  All spectra exhibit abundances typical of those in the ISM with a couple instances of depleted Si and Mg and one instance of a loosely constrained Si overabundance.  Overall, these fits are consistent with varying equilibrium conditions and plasma temperatures in an ISM dominated medium.

\section{Discussion}
\subsection{Shell Emission}
These data are the first to show a clear shell of X-ray emission at the limb of SNR G272.2-3.2.  As previously noted, based on their measured spectral index \citet{Duncan1997} argue for a shell type morphology, which is concurrent with our findings.  The morphology within the X-ray shell is dominated by shock heated ISM located in clumps and filaments along the rim of the remnant as seen in the false color image and the spectral fits.  Calculations of shock velocity and density support this conclusion.  Solving the Rankine-Hugoniot relations in the strong shock case with $\gamma = 5/3$ gives the post-shock temperature as a function of shock velocity, $kT=(3/16) \mu m_{p} v^2$, with $\mu = 0.6$ for an ionized plasma.  We assume that $T_e\sim T_{ion}$ as expected through Coulomb collisions at a temperature of $\sim10^6$ K and $n_e\sim1$ cm$^{-3}$.  Such a plasma can equilibrate within a few hundred years or even more rapidly given higher densities (see eqn. 36.37 in \citet{Draine}).  If the electrons and ions are not in temperature equilibrium then the calculated velocities signify lower bounds.  The calculated velocities for the lower temperature equilibrium components range from 370--490 km sec$^{-1}$ with velocities of 620--970 km sec$^{-1}$ for the higher temperature nonequilibrium regions.  These shock speeds are significantly decelerated from the initial blast wave velocity of $>5000$ km/s.  Interactions with a clumpy ISM could produce such results with slow shock propagation speeds and rapid equilibration in higher density clouds interspersed with higher shock velocities, higher temperatures, and nonequilibrium conditions in more rarefied regions.

Densities in the shell are calculated using the \textit{norm} parameter where $norm=[10^{-14}/(4 \pi D^2)] \int n_e n_H dV$. The integral contains the emission measure for the plasma which is dependent on the density and total emitting volume.  Following the considerations taken in the \citet{Harrus} study, we adopt 5 kpc for the distance to G272.2-3.2.  We also assume that $n_e=1.2n_H$.  In regions G, H, J, L, N, and O we calculate the volume using a rectangle with the same area as the region with a depth equal to the long dimension of that rectangle.  For the remaining regions we use the volume of a sphere with a radius that approximates the size of the extraction region.  We also include a filling factor, $f$, such that $V=4/3 \pi R^3 f$, for example.  In the equilibrium regions of A, E, F, G, and J we find relatively high electron densities of 1.0, 3.9, 1.5, 0.5, and 1.5 cm$^{-3}$ $D_5^{-1/2}$ $f^{-1/2}$, respectively, where $D_5$ is the distance in units of 5 kpc.  The nonequilibrium regions are more rarefied with densities ranging from 0.13--1.1 cm$^{-3}$ $D_5^{-1/2}$ $f^{-1/2}$ giving an average density of 0.46 cm$^{-3}$ $D_5^{-1/2}$ $f^{-1/2}$.  These calculations are consistent with a conclusion of slower velocities and equilibrium conditions in the denser media, again signifying interactions in an anisotropic, clumpy shell.

We can also estimate the age of the remnant from the equilibrium shock velocities.  Assuming that the remnant can be described by a Sedov solution \citep{Spitzer}, the time since shock heating can be calculated as $t=(2/5)R/v$, where $R$ is the estimated shock distance from the center.  Measurements of the diameter of the X-ray shell emission average $\sim$15 arcminutes, giving a radius of 11 pc $D_5$.  The calculated ages range from 8800--12000 years $D_5$, consistent with the $\sim$6000--15000 years calculation from \citet{Harrus} and considerably older than the $\sim$6000 year upper limit in previous studies.  The discrepancy with the latter is most likely due to the assumption of a Sedov type evolution of shock propagation into an isotropic, low density medium.  The presence of the ISM clouds in a shell, either swept up by the SN explosion or already present in a precursor cavity created by the progenitor stellar wind, lead to rapid shock deceleration and a possible overestimate of the age.  Alternatively, we can use the ionization time scale, $\tau=n_et$ to determine the time since shock heating of the nonequilibrium plasmas in the shell.  When considering regions with well constrained ionization parameters (not consistent with zero at the $>4\sigma$ level; B, H, I, L, and N) the ionization times range from 3600--10000 years $D_5^{1/2}$ $f^{1/2}$ with an average of 6300 years $D_5^{1/2}$ $f^{1/2}$, which is probably more representative of the age of the shell emission.

\begin{figure} [htbp]
   \centering
%   \epsscale{1.0}
%   \plotone{f6.eps}
%   \includegraphics[width=6.0in,height=3.0in]{figures/ir.pdf}
   \includegraphics[width=6.0in,height=3.0in]{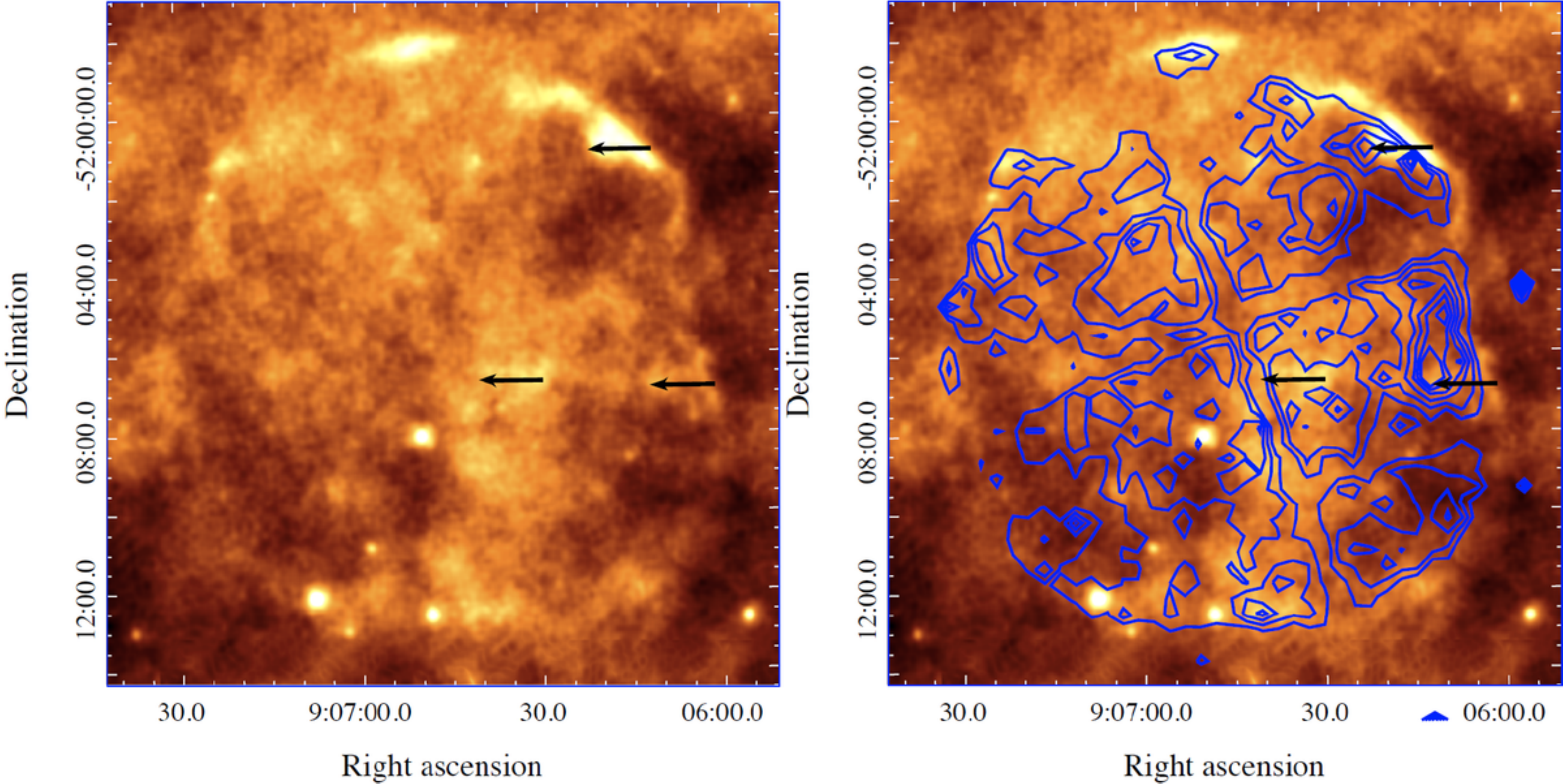}
   \caption{\textit{Left}--The \textit{WISE} raw infrared data in the 24$\mu$m band.  The locations of optical filaments studied by \citet{Winkler1993} are marked with black arrows.  \textit{Right}--The \textit{WISE} infrared image with X-ray contours in blue.  There are seven X-ray contour levels evenly spaced between 18--42 counts per pixel.}
   \label{ir}
\end{figure}

\citet{Duncan1997} and \citet{Winkler1993} find clumpy radio emission and optical filaments within G272.2-3.2.  Their conclusion of interactions with anisotropic circumstellar material is consistent with our findings in the X-ray shell.  The remnant is quite dim in the radio so a lack of sensitivity or an unoptimal array configuration could result in the emission concentrated into blobs as opposed to outlining a more complete shell as we see in the X-ray.  To probe deeper into the circumstellar environment of G272.2-3.2 we utilize \textit{WISE} infrared, 24$\mu$m images.  Interactions with dense ISM clouds should result in continuum emission from shock heated dust.  The infrared image is shown in Figure \ref{ir}.  An overlay of X-ray contours is shown on the right of this figure.  As in X-rays, this image readily displays a nearly complete circular, swept of shell of ISM including bright shocked filaments.  In fact, aside from region O, every feature that we analyze in the X-ray is strongly correlated to an IR cloud in the shell.  There is also lower brightness emission filling much of the interior of the remnant.  Some of this may be located in the SNR shell along the line of sight, but much of the emission may not result from G272.2-3.2 at all given the complex infrared environment in the vicinity.  

Black arrows are drawn on Figure \ref{ir} to denote the position of the optical filaments studied by \citet{Winkler1993} and radio correlated by \citet{Duncan1997}.  Evident infrared blobs are also present in these positions, most notably in the northwest rim near a bright emission feature.  The innermost optical position falls along the chip spacing in the X-ray rendering a correlation impossible.  However, the optical filaments in the west and northwest are well correlated with X-ray knots in the shell, in contradiction to the lower resolution \textit{ROSAT} correlation performed in \citet{Duncan1997}.  These regions are therefore also correlated to the radio emission detected in \citet{Duncan1997}.  In fact, the brightest radio feature corresponds to the brightest X-ray knot in the shell and the location of the western optical filaments.  These X-ray data compliment the \citet{Duncan1997} radio findings well and corroborate their conclusion of a shell type remnant.  

The western region is also intriguing because of its X-ray morphology.  The nearly circular shape of G272.2-3.2 is noticeably indented on the western edge.  It appears as if the shock wave that has swept up the shell is running into a large and particularly dense cloud in the environment, thus leading to the brightest sites of emission in the X-ray, optical, and radio lying along this front.  Our region J is located at the very edge of the X-ray emission along the area that appears the most indented.  The emission is noticeably red, cool, and dense as expected from an interaction with such a cloud.  The K region located just to the interior is the location of the brightest X-ray knot and the optical filament.  This region is hotter, more rarefied and not in equilibrium.  The X-ray plasma could have been more recently shock heated by a reverse shock propagating inward from the shock/cloud interaction in J.  The strong radio emission in this area, while previously correlated to the optical emission in K, must lie nearer to the edge in region J.  The centers of these two regions are only separated by 50 arcseconds which is equivalent to the beam sizes used in \citet{Duncan1997}.  Therefore, it is reasonable to conclude that this interaction site consists of a large shock/cloud interaction giving rise to emission in all bands.  The exact location of the emission however depends on the densities in the clumpy ISM: swept up magnetic fields lead to radio synchrotron at the lowest densities; increased density results in collisional ionization to produce X-ray emission; increasing it further gives slow shocks and radiative cooling in optical filaments; while the highest densities show blackbody emission from shock heated dust. 

\subsection{Interior Emission}
Previous X-ray studies \citep{SezerGok,Harrus} utilize data from lower resolution telescopes.  This made it difficult to discern the knotty, lower energy emission near the limb from the hotter material just interior to it.  The evident X-ray morphology in these studies was centrally peaked with the lack of a shell.  To explain this morphology the studies investigate two different thermal models, a cloud evaporation model \citep{WhiteLong} and a thermal conduction model \citep{Cox1999}.  The former models clouds evaporating in a hot shocked medium while the latter models temperature equilibration in a Sedov remnant due to thermal conduction.  Both studies side with the cloud evaporation model and agree that available detail on the interior emission is poor.  Now however, the high resolution of \textit{Chandra} clearly shows that the morphology actually is not centrally peaked at all and that the physical situation is different than that assumed by these models.  The highest brightness regions in the \textit{Chandra} data are found in discrete filaments at the limb in a shell.  These features are noticeably spatially separated from the hot, diffuse interior emission.  The previous lower resolution observations blurred the discrete, bright regions into the dark annulus, just to the interior.  This ``diffusion'' of the shell flux decreases its significance and falsely results in a centrally peaked morphology since the central emission is truly diffuse.  

The physical characteristics suggested by \textit{Chandra} are a hot, ejecta dominated interior and a cooler shell of shocked ISM, which only appears near the limb where the emitting column is sufficiently deep.  The low energy flux is dominated by the interaction at the limb with the highest overall brightnesses in discrete features.  Toward the center (the inner $\sim$8 arcminutes), the line of sight probes into the central emission which is filled with hot plasma, but not the peak of emission.  There does not appear to be a significant contribution from shell emission in these central regions, i.e. confusion is minimal along the line of sight.  It is possible that cloud evaporation or thermal conduction is still important in the interior.  The fits in the center region and annulus 1 (the central 5.2 arcminutes) are consistent with a constant temperature in the diffuse interior as expected in both models, however the thermal conduction model requires equilibrium conditions which are clearly not present.  Annulus 2 has a component that is consistent with this temperature as well but is confused by contributions from a second component which may or may not be due to the dark annulus.  

The densities found in these regions are fairly consistent as well.  We use the volume of a sphere for the center region and for annulus 1 we use the volume of an annular cylinder with line of sight depth equal to the outer radius of the annulus.  The calculated densities are 0.10 cm$^{-3}$ $D_5^{-1/2}$ $f^{-1/2}$ in both regions.  The time since shock heating for these nonequilibrium plasmas is comparable at 8500 and 8900 years $D_5^{1/2}$ $f^{1/2}$.  Given the strong shell morphology and the hot interior plasma it seems likely that the hot, nonequilibrium interior conditions are caused by a reverse shock orginating from the ISM interaction at the SNR limb.  This is similar to the morphology seen in other thermal composite SNR such as Kes 27(Yang Chen et al. 2008 ApJ 676 1040).  However, if this was the case then the ionization timescale should be much smaller, not larger than that found in the shell.  Using an age of 8700 years at the center and 6300 years at the limb located 11 pc away suggests a shock speed of 4500 km sec$^{-1}$ if both plasmas were created by the same shock, reasonable for a supernova explosion.  Therefore, instead of a reverse shock scenario, it appears that the interior has only been shocked once and at a time prior to interactions in the shell.  The constant temperature, constant density, and age evolution support ongoing equilibration of the shock heated, ejecta dominated plasma as the cause of the central, diffuse morphology.  Furthermore, this suggests that the most accurate age estimate of G272.2-3.2 should be consistent with the ionization timescale, $\sim$8700 years.

The enriched Fe ejecta suggest a Type Ia origin for G272.2-3.2.  Based on a study of its morphology, \citet{Lopez} agree with this classification, while \citet{SezerGok} find similar overabundances and conclude that a Type Ia progenitor is most likely.  We perform a point source analysis on the remnant, and while we find several regions that are consistent with a \textit{Chandra} psf, none have significant flux to spectrally verify a compact object origin.  Furthermore, there are no morphologies surrounding these sources that suggest a possible pulsar such as a PWN or associated bow shocks, cometary tails, etc.  Our spectral fits find no contributions from nonthermal components anywhere in this remnant.  The radio study by \citet{Duncan1997} also concludes that there is no evidence of plerionic emission in this object.  All of these findings support a Type Ia classification.

\begin{figure} [htbp]
   \centering
%   \epsscale{1.0}
%   \plotone{f7.eps}
%   \includegraphics[width=6.0in,height=4.8in]{figures/nomoto.pdf}
   \includegraphics[width=6.0in,height=4.8in]{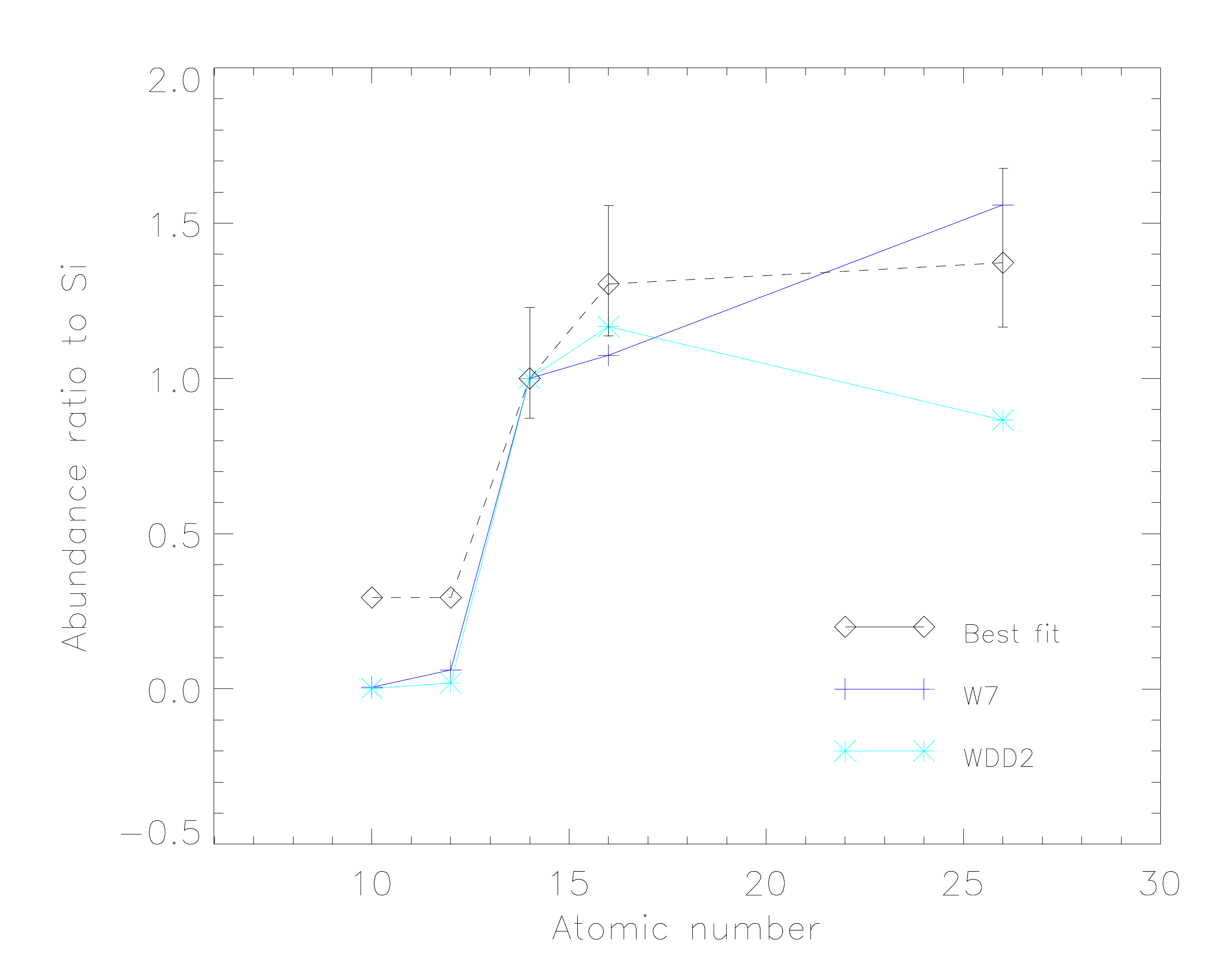}
   \caption{Nucleosynthesis yields of Type Ia supernovae based on the \citet{Nomoto} deflagration (W7; blue crosses) and delayed detonation (WDD2; cyan asterisks) models.  The abudances from the best fit models are shown as black diamonds for comparison.}
   \label{nomoto}
\end{figure}

Following the analysis performed by \citet{SezerGok}, we consider the overabundances found in our fits to determine if they are consistent with the \citet{Nomoto} models of a Type Ia supernova.  Figure \ref{nomoto} shows the theoretical abundances of ejecta based on nucleosynthesis yields from a carbon deflagration explosion model (W7 in \citet{Nomoto}) and a delayed detonation model (WDD2 in \citet{Nomoto}).  We also plot our best fit values for Ne, Mg, Si, S, and Fe.  The ratios of S/Si and Fe/Si that we find are consistent with these Type Ia nucleosynthesis yields, further suggesting a Type Ia classification for G272.2-3.2.  While the S ratio is more in-line with the delayed detonation model, our Fe value is much closer to the typical deflagration model.  The abundances for Ne and Mg were held constant at solar levels since their variability did not statistically improve the fits, hence the lack of error bars in this figure.  The solar level for these two species, relative to higher atomic number elements, is larger than that expected in a Type Ia supernova.  The presence of higher levels of Ne and Mg in the fits could be due to poor statistics in the spectra or line of sight contributions from plasma in the shell.  In fact, inspection of the annular spectra in Figure \ref{annularspecs} show that the plasma at the center and annulus 1 (and likewise the blue band of the RGB image) are dominated by the strong lines from the He-like species of \ion{Si}{13} at $\sim$1.8 keV and \ion{S}{15} at $\sim$2.5 keV.  The lower energy emission in these regions is dominated by the complex of Fe L-shell emission lines between 0.8--1.2 keV, which is evident by the red band visible in the interior.  However, as we move outward toward annulus 4 the overabundances of these species decreases while we see an increase in the contribution of ISM plasma.  This is evident as a relative increase in \ion{Ne}{9} at $\sim$0.9 keV, \ion{Ne}{10} at $\sim$1.0 keV, and \ion{Mg}{11} at $\sim$1.3 keV in comparison to Si, S, and Fe lines until solar levels are reached.  The strong Ne and Mg lines in the shell could be contributing to the flux in the central regions.

\section{Summary}
G272.2-3.2 exhibits significant X-ray emission and through the use of the superior spatial resolution of \textit{Chandra} we see that it is not as centrally peaked as previously thought.  There is a clear shell of X-ray emission that is well characterized by shock heated ISM.  There are filaments and knots around the entire rim with a range of equilibrium conditions, temperature, shock speed, and density suggesting that the shell is clumpy and far from isotropic.  There are likely radio and optical correlations to the brightest X-ray region in the west which also signifies interaction with a large ISM cloud.  Investigation of infrared images from \textit{WISE} show a clear shell as well that is strongly correlated to the emission in the X-ray shell.  

Much of the high energy emission is in fact located toward the center in a diffuse internal region that appears spatially separated from the shell.  It is characterized by nonequilbrium conditions in a hot plasma, $kT\sim1.5$ keV, with overabundant Si, S, and Fe.  The relative abundances follow expected nucleosynthesis yields from Type Ia supernovae.  Furthermore, the lack of both plerionic emission and likely point sources point toward a Type Ia origin.  An ionization age of 8700 years at the center with an average ionization age of 6300 years in the shell suggest a pleasant picture of a supernova shock originating 8700 years ago, shock heating lower density ($\sim$0.1 cm$^{-3}$) ejecta near the center, then propagating 11 pc at $\sim$4500 km sec$^{-1}$ to shock heat swept up ISM in a shell.  Such a scenario would argue for a 8700 year age for G272.2-3.2.

\section{Acknowledgments}\label{ack}
The authors would like to acknowledge internal funding initiatives at the University of Iowa for support of this work.  Thomas Brantseg is supported by NASA grant NNX10AN16H.  The X-ray data used here were obtained from the Chandra Data Archive.  In addition, this publication makes use of data products from the Wide-field Infrared Survey Explorer, which is a joint project of the University of California, Los Angeles, and the Jet Propulsion Laboratory/California Institute of Technology, funded by the National Aeronautics and Space Administration.

\clearpage

%Table 1:

\begin{deluxetable} {c c c c c c c c c c c}
 \tablewidth{8 in}
 %\scalebox{0.5}{%
 \tabletypesize{\scriptsize}
 \rotate
 \tablecaption{Best fit parameters for annular spectral extraction regions}
 \tablehead{
  \colhead{Region} &
  \colhead{$\chi ^2$/$\nu$} &
  \colhead{$kT_{vnei1}$} &
  \colhead{$kT_{vnei2}$} &
  \colhead{$kT_{vapec}$} &
  \colhead{$\tau1$} &
  \colhead{$\tau2$} &
  \colhead{$norm_{vnei1}$} &
  \colhead{$norm_{vnei2}$} &
  \colhead{$norm_{vapec}$} &
  \colhead{$N_H$} \\
  & & (keV) & (keV) & (keV) & ($10^{10}$ s cm$^{-3}$) & ($10^{10}$ s cm$^{-3}$) & $10^{-3}A$\tablenotemark{a} & $10^{-3}A$\tablenotemark{a} & $10^{-3}A$\tablenotemark{a} & ($10^{22}$ cm$^{-2}$)
 }
 \startdata
 Center & 82/89 & $1.50^{+0.10}_{-0.06}$ & \nodata & \nodata & $2.7^{+0.2}_{-0.3}$ & \nodata & $0.12^{+0.03}_{-0.04}$ & \nodata & \nodata & $1.07 \pm 0.04$ \\
 Annulus 1 & 183/143 & $1.43 \pm 0.03$ & \nodata & \nodata & $2.8 \pm 0.2$ & \nodata & $0.53 \pm 0.06$ & \nodata & \nodata & $1.07 \pm 0.02$ \\
 Annulus 2 & 250/160  & $1.4^{+0.3}_{-0.1}$ & $0.9^{+0.1}_{-0.7}$ & \nodata & $6.2^{+0.9}_{-2.3}$ & $3.8_{-0.3}$ & $0.35^{+0.07}_{-0.11}$ & $0.13^{+0.39}_{-0.03}$ & \nodata & $1.13^{+0.02}_{-0.19}$ \\
 Annulus 3 & 169/169 & $0.72^{+0.03}_{-0.04}$ & \nodata & \nodata & $8.8^{+1.8}_{-0.8}$ & \nodata & $3.2^{+0.3}_{-0.1}$ & \nodata & \nodata & $1.06^{+0.03}_{-0.02}$ \\
 Annulus 4 & 175/176 & $1.08^{+0.14}_{-0.04}$ & \nodata & $0.226^{+0.008}_{-0.006}$ & $3.6 \pm 0.2$ & \nodata & $4.1^{+0.3}_{-0.7}$ & \nodata & $83^{+16}_{-13}$ & $1.03 \pm 0.02$ \enddata
 \tablenotetext{a}{Normalization parameter, where $A=[10^{-14}/(4\pi D^2)]  \int n_e n_H dV$. D is the distance to the LMC and the integral is the volume emission measure.}
 \label{table1}
\end{deluxetable}

\clearpage

%Table 2

\begin{deluxetable} {c c c c c c c c c c c}
 \tablewidth{5.6 in}
 %\scalebox{0.5}{%
 \tabletypesize{\scriptsize}
 \rotate
 \tablecaption{Elemental abundances for annular spectral extraction regions}
 \tablehead{
  \colhead{Regions} &
  \colhead{O} &
  \colhead{$\sigma$} &
  \colhead{Si} &
  \colhead{$\sigma$} &
  \colhead{S} &
  \colhead{$\sigma$} &
  \colhead{Fe} &
  \colhead{$\sigma$ }&
  \colhead{Mg} &
  \colhead{$\sigma$} \\
 }
 \startdata
 Center & $14^{+11}_{-5}$ & 3.3 & $4.2^{+1.6}_{-0.8}$ & 4 & $4.0^{+1.3}_{-0.9}$ & 3.3 & $5^{+2}_{-1}$ & 4 & \nodata & \nodata \\
 Annulus 1 & $9^{+3}_{-2}$ & 4 & $2.9 \pm 0.3$ & 6.3 & $4.8 \pm 0.5$ & 7.6 & $4.3^{+0.7}_{-0.6}$ & 5.5 & \nodata & \nodata \\
 Annulus 2 & $9.0^{+0.9}_{-3.1}$ & 2.6 & $3.1^{+0.2}_{-0.4}$ & 5.3 & $4.5^{+0.3}_{-0.6}$ & 5.8 & $4.7^{+0.3}_{-1.2}$ & 3.1 & \nodata & \nodata \\
 Annulus 3 & $5.9^{+0.5}_{-0.8}$ & 6.1 & $1.6 \pm 0.1$ & 6 & $3.1^{+0.3}_{-0.4}$ & 5.3 & $2.1 \pm 0.2$ & 5.5 & \nodata & \nodata \\
 Annulus 4 & $1.5 \pm 0.1$ & 5 & \nodata & \nodata & \nodata & \nodata & \nodata & \nodata & $0.80 \pm 0.03$ & 6.7 \enddata
 \tablecomments{Abundances are given with 1$\sigma$ errors and relative to solar, with solar values = 1.  Only variable abundances are shown with the rest held at solar values.  The $\sigma$ columns represent deviation above or below solar abundance for each element.}
 \label{table2}
\end{deluxetable}

\clearpage

%Table 3:

\begin{deluxetable} {c c c c c c c c c c}
 \tablewidth{7 in}
 \tabletypesize{\scriptsize}
 \rotate
 \tablecaption{Best fit parameters for outer spectral extraction regions}
 \tablehead{
  \colhead{Region} &
  \colhead{$\chi ^2$/$\nu$} &
  \colhead{$kT_{vnei}$} &
  \colhead{$kT_{vapec}$} &
  \colhead{$\tau$} &
  \colhead{$norm_{vnei}$} &
  \colhead{$norm_{vapec}$} &
  \colhead{$N_H$} &
  \colhead{Si} &
  \colhead{Mg} \\
  & & (keV) & (keV) & ($10^{10}$ s cm$^{-3}$) & $10^{-3}A$\tablenotemark{a} & $10^{-3}A$\tablenotemark{a} & ($10^{22}$ cm$^{-2}$)
 }
 \startdata
 A & 27/34 & \nodata & $0.28^{+0.02}_{-0.04}$ & \nodata & \nodata & $2.2^{+2.1}_{-0.5}$ &  $1.14^{+0.08}_{-0.06}$ & \nodata & \nodata \\
 B & 49/39 & $0.63^{+0.34}_{-0.06}$ & \nodata & $7^{+3}_{-1}$ & $0.23^{+0.08}_{-0.21}$ & \nodata & $0.92^{+0.06}_{-0.02}$ & \nodata & \nodata \\
 C & 48/56 & $0.57^{+0.17}_{-0.03}$ & \nodata & $14^{+3}_{-4}$ & $0.51^{+0.21}_{-0.06}$ & \nodata & $0.85^{+0.03}_{-0.07}$ & \nodata & \nodata \\
 D & 79/56 & $0.59^{+0.03}_{-0.05}$ &\nodata & $70^{+70}_{-30}$ & $0.50^{+0.10}_{-0.06}$ & \nodata & $0.98^{+0.06}_{-0.05}$ & \nodata & \nodata \\
 E & 53/39 & \nodata & $0.16^{+0.02}_{-0.01}$ & \nodata & \nodata & $44^{+52}_{-22}$ & $1.43^{+0.11}_{-0.09}$ & $10^{+6}_{-3}$ & \nodata \\
 F & 49/48 & \nodata & $0.25^{+0.04}_{-0.02}$ & \nodata & \nodata & $5 \pm 2$ & $1.20^{+0.05}_{-0.08}$ & \nodata & \nodata \\
 G & 43/49 & \nodata & $0.62 \pm 0.03$ & \nodata & \nodata & $0.40 \pm 0.04$ & $0.87 \pm 0.03$ & \nodata & \nodata \\
 H & 31/38 & $0.45^{+0.04}_{-0.16}$ & \nodata & $12^{+17}_{-3}$ & $0.8^{+3.0}_{-0.2}$ & \nodata & $1.06^{+0.20}_{-0.05}$ & \nodata & $0.8^{+0.1}_{-0.3}$ \\
 I & 77/68 & $0.66^{+0.22}_{-0.02}$ & \nodata & $8 \pm 1$ & $0.52^{+0.04}_{-0.21}$ & \nodata & $0.87^{+0.04}_{-0.01}$ & \nodata & \nodata \\
 J & 42/36 & \nodata & $0.23^{+0.02}_{-0.01}$ & \nodata & \nodata & $2.1^{+0.8}_{-0.7}$ & $0.87 \pm 0.05$ & \nodata & \nodata \\
 K & 44/41 & $0.58 \pm 0.03$ & \nodata & $70^{+50}_{-20}$ & $0.29 \pm 0.03$ & \nodata & $0.69 \pm 0.03$ & \nodata & \nodata \\
 L & 71/68 & $1.1^{+0.2}_{-0.1}$ & \nodata & $2.7^{+0.3}_{-0.4}$ & $0.21^{+0.04}_{-0.05}$ & \nodata & $0.72^{+0.03}_{-0.05}$ & $0.6 \pm 0.1$ & \nodata \\
 M & 51/45 & $0.9 \pm 0.3$ & \nodata & $3.6^{+0.9}_{-1.2}$ & $0.20^{+0.12}_{-0.08}$ & \nodata & $0.8^{+0.1}_{-0.2}$ & \nodata & \nodata \\
 N & 116/91 & $0.8^{+0.3}_{-0.2}$ & \nodata & $4.0^{+3.0}_{-0.7}$ & $0.7^{+0.5}_{-0.3}$ & \nodata & $0.91^{+0.09}_{-0.06}$ & \nodata & $0.76^{+0.06}_{-0.08}$ \\
 O & 27/25 & $1.1 \pm 0.3$ & \nodata & $2.4 \pm 0.7$ & $0.11^{+0.06}_{-0.03}$ & \nodata & $1.01^{+0.08}_{-0.09}$ & $0.4^{+0.2}_{-0.1}$ & \nodata \enddata
 \tablecomments{Abundances are given relative to solar, with solar values = 1.  Only variable abundances are shown with the rest held at LMC values.}
 \tablenotetext{a}{Normalization parameter, where $A=[10^{-14}/(4\pi D^2)]  \int n_e n_H dV$. D is the distance to the LMC and the integral is the volume emission measure.}
 \label{table3}
\end{deluxetable}

\clearpage

%figure 1:

\clearpage

%figure 2:

\clearpage

%figure 3:

\clearpage

%figure 4:

\clearpage

%figure 5:

\clearpage

%figure 6:

\clearpage

%figure 7:

\clearpage

\end{document}